
\input harvmac
\Title{FERMI-PUB-91/338-T}{Perturbations of a Stringy Black Hole}
\centerline{{\bf E. Raiten}\footnote{$^\dagger$}
{e-mail:\ Raiten@FNAL}}
\bigskip\centerline{Theory Group, MS106}
\centerline{Fermi National Accelerator Laboratory}
\centerline{P.O. Box 500, Batavia, IL 60510}

\vskip .3in
We extend the three dimensional stringy black hole of Horne and
Horowitz to four dimensions.  After a brief
discussion of the global properties of the metric, we discuss
the stability of the background with respect to
small perturbations, following the methods of
Gilbert and of Chandrasekhar.  The potential for axial perturbations is found
to be positive definite.

\Date{12/91}

%

\def\np#1{{\it Nucl. Phys.} {\bf B#1}}

\def\prd#1{{\it Phys. Rev.}{\bf D#1}}

\def\aj#1{{\it Astrophys. Journal}{\bf #1}}
\def\fa{e^{-2f_0}}
\def\fb{e^{-2f_1}}
\def\fc{e^{-2f_2}}
\def\fd{e^{-2f_3}}
\def\fap{e^{2f_0}}
\def\fbp{e^{2f_1}}
\def\fcp{e^{2f_2}}
\def\fdp{e^{2f_3}}
\def\ca{\chi_1}
\def\cb{\chi_2}
\def\cc{\chi_0}
\def\cd{\chi_3}

\nref\garf{D. Garginkle, G. Horowitz, A. Strominger, \prd{43} (1991) 3140.}
\nref\triv{A. Shapere, S. Trivedi, F. Wilczek, \lq\lq Dual Dilaton Dyons''
IASSNS-HEP-91/33, June 1991.}
\nref\gnga{G. Gilbert, \lq\lq On the
Perturbations of String Theoretic Black Holes'',
Maryland preprint UMDEPP 92-035, Aug. 1991.}
\nref\gngb{G. Gilbert,
\lq\lq The Instability of String-Theoretic Black Holes'',
Maryland preprint UMDEPP 92-110, Nov. 1991.}
\nref\witta{E. Witten, \prd{44} (1991) 314.}
\nref\barsa{I. Bars, \lq\lq String Propagation on Black Holes, USC
preprint USC-91/HEP-83, May 1991.}
\nref\barsb{I. Bars, K. Sfetsos, \lq\lq A Superstring in Four Curved
Space-Time Dimensions'' USC preprint USC-91/HEP-B6, Nov. 1991.}
\nref\hh{J. Horne, G. Horowitz, \lq\lq Exact Black String Solutions
in Three Dimensions'', UCSB preprint UCSBTH-91-39, Jul. 1991.}
\nref\hor{P. Ho\c rava, \lq\lq Some Exact Solutions of String Theory
in Four and Five Dimensions'', Chicago preprint EFI-91-57, Oct. 1991.}
\nref\callan{C. Callen, R. Myers, M. Perry, \np{311} (1988) 673.}
\nref\egu{T. Eguchi, \lq \lq Topological Field Theory and the Space-Time
Singularity'', EFI 91-58, Oct. 1991.}
\nref\chandraa{S. Chandrasakhar, J. Friedman, \aj{175} (1972) 379.}
\nref\chandrab{S. Chandrasekhar, J. Hartle, {\it Proc. Roy. Soc. Lond.}
{\bf A384}, (1982) 301.}
\nref\ejra{E. Raiten, in progress.}
\nref\gngc{G. Gilbert, private communication.}
\nref\dvv{R. Dijkgraaf, H. Verlinde, E. Verlinde, \lq\lq String Propagation
in a Black Hole Geometry'', PUPT-1252, IASSNS-HEP-91/22, May 1991.}
\nref\lyk{S. Chaudhuri, J. Lykken, in preparation.}
\nref\ellis{J. Ellis, N. Mavromatos, D. Nanopoulos, \lq\lq On the
Evaporation of Black Holes in String Theory'', CERN-TH.6309/9.6309/91, 91.}
\vfill
\eject

\newsec{Introduction}
The discovery of exact and approximate solutions of perturbative
string theory with black hole behavior has garnered much attention
recently.  In addition to furthering our understanding of the nature
of the solutions of the field equations of general relativity and/or
the nonperturbative content of string theory, one might
even hope that a stringy black hole would have observational
and measureable features which might not require a detailed knowledge
of how the fundamental string theory devolves into the Standard Model.

In one approach\garf \triv , one starts with a
so-called \lq\lq string inspired'' low energy effective
action that would naturally devolve from a fundamental string
theory.  The key difference from a non-string inspired model
is then the presence of the dilaton.

The presence of the dilaton has important effects.  Gilbert has
recently shown that the four dimensional charged black hole derived
in \garf\ is catastrophically unstable to small perturbations, in
sharp contrast to similar backgrounds in general relativity\refs{\gnga,\gngb}.
This instability is completely distinct from other effects, such
as possible quantum mechanical evaporation.  It may be difficult,
however, to further investigate the \lq\lq stringiness'' of these solutions,
as it is not known what conformal field theory they correspond to,
and such solutions generically include an electromagnetic field
whose stringy origins are somewhat unclear, while they often do not
contain an axion field which is expected on general grounds.

In the second approach, gauged Wess-Zumino-Witten (WZW) models
have been found to correspond to string propagation in curved
backgrounds, backgrounds with black hole singularities with
an appropriate choice of groups and gauged subgroups.  Higher dimension
analogs of Witten's two dimensional black hole\witta have been
found by Bars and Sfetsos\refs{\barsa,\barsb}, Horne and Horowitz \hh
and Horava \hor .  The availability of such higher dimensional
analogs, especially in four dimensions, presents some potentially
important oppourtunities.
It would be very interesting, for example, to consider a four
dimensional black hole of this type and subject it to the same
perturbative analysis which the \lq\lq string-inspired'' model has
recently undergone.  That is precisely what will be attempted
here.

First, in Section 2,
we will extend the black hole of Horne and Horowitz \hh\ to
four dimensions.  The global properties
of the metric will be briefly discussed
in Section 3.  The metric, dilaton, and axion have sufficiently
simply forms that the perturbative analysis in Section 4 will not be overly
difficult.
The equations for the perturbations separate, as
in \gnga , into two sets, referred to as axial and polar.  The
potential for the axial perturbations will be explicitly derived.
Numerical evidence strongly suggests that this potential is strictly
positive, although this is not obvious from the analytic
expressions.   Further discussion and conclusions will be
collected in Section 5.
\vfill
\eject

\newsec{A Four Dimensional Black Hole}
Let us begin by reveiwing the three dimensional black string of
Horne and Horowitz\hh , which in turn is an extension of Witten's
2-d black hole based on the gauged $SL(2,R)/U(1)$ WZW model.
We begin with a Lorentz metric $ds^2=2d\sigma_+
\sigma_-$ on the world sheet $\Sigma$.
In terms of the group valued function $g$, the WZW action is
\eqn\wzw{L(g)={k\over {4\pi}}\int d^2\sigma Tr(g^{-1}\del_+gg^{-1}\del_-g)-
{k\over {12\pi}}\Gamma (g).}
$\Gamma (g)$ is the Wess-Zumino-Witten term, normally written as an
integral over a three manifold $B$ whose boundary is $\Sigma$,
\eqn\gam{\Gamma(g)=\int_Bd^3yTr(g^{-1}dg\wedge g^{-1}dg\wedge g^{-1}dg).}

We now gauge a one dimensional subgroup $H$ of the symmetry group, with
action $g\rightarrow hgh$.  The symmetry is promoted to a local symmetry
by the introduction of a gauge field $A_i$, with values in the Lie algebra
of $H$.  The local symmetry acts as
\eqn\symm{\delta g=\epsilon g+g\epsilon, \delta A_i=-\del_i\epsilon ,}
where $\epsilon$ is an infinitesimal parameter.  The action for which
this transformation law is a symmetry, the gauged WZW action, is
\eqn\gwzw{L(g,A)=L(g)+{k\over {2\pi}}\int d^2\sigma Tr(A_+\del_-gg^{-1}+
A_-g^{-1}\del_+g+A_+A_-+A_+gA_-g^{-1}).}
The 2-d black hole of Witten\witta\ follows by letting $G=SL(2,R)$, with $H$
the subgroup generated by $\pmatrix{1&0\cr 0&-1}$,
and integrating out the gauge fields,
which appear quadratically in the action.

The extension of Horne and Horowitz\hh follows by adding a free boson $x_1$ to
the action, that is, letting $G=SL(2,R)\times R$, and by simultaneously
tensoring in translations in $x_1$ to the subgroup $H$.
The extension we will consider here is to add another free boson
$x_2$, and mod
out by translations in both $x_1$ and $x_2$.  Parametrizing $SL(2,R)$
as
\eqn\slr{g=\pmatrix{a&u\cr -v&b}}
we have the ungauged action
\eqn\acta{\eqalign{L(g)=- &{k\over {4\pi}}\int d^2\sigma (\del_+u\del_-v+
\del_-u\del_+v+\del_+a\del_-b+\del_-a\del_+b)\cr + &{k\over {2\pi}}
\int d^2\sigma logu(\del_+a\del_-b-\del_-a\del_+b)+{1\over {\pi}}
\int d^2\sigma \Sigma_i\del_+x_i\del_-x_i.}}
The gauge transformations are
\eqn\gt{\delta a=2\epsilon a, \delta b=-2\epsilon b, \delta u=\delta v=0,
\delta x_i=2\epsilon c_i, \delta A_i=-\del_i\epsilon ,}
where the $c_i$ are constants.  The gauged action is then
\eqn\actb{L(g,A)=L(g)\eqalign{&{k\over {2\pi}}\int d^2\sigma
A_+(b\del_-a-a\del_-b
-u\del_-v+v\del_-u+{{4c_i}\over k}\del_-x_i)\cr + &
{k\over{2\pi}}\int{d^2}\sigma A_-(b\del_+a-a\del_+b
+u\del_+v-v\del_+u+{{4c_i}\over k} \del_+x_i)\cr + &
{k\over{2\pi}}\int{d^2}\sigma 4A_+A_-(1+{{2c^2}\over k}
-uv),} }
where a sum on $i=1,2$ is assumed, and $c^2=c_1^2+c_2^2$.  We now fix
the gauge by setting $a=\pm b$, depending on the sign of $1-uv$, and
integrate out the gauge fields, thus obtaining
\eqn\actsum{L=L_1+L_2+L_3,}
\eqn\La{L_1=-{k\over {8\pi}}\int d^2\sigma {{\lambda_t (v^2\del_+u\del_-u
+u^2\del_+v\del_-v)+(2-2uv+2\lambda_t-\lambda_tuv)(\del_+u\del_-v+
\del_-u\del_+v)}\over {(1-uv)(1+\lambda_t-uv)}}}

\eqn\Lb{L_2={1\over {\pi}}\int d^2\sigma {{1+\lambda_2-uv}\over
{1+\lambda_t-uv}}\del_+x_1\del_-x_1+(1\leftrightarrow 2)-
{{(\lambda_1\lambda_2)^{1/2}}\over {1-\lambda_t-uv}}(\del_+x_1\del_-x_2+
\del_-x_1\del_+x_2),}
\eqn\Lc{L_3={1\over {2\pi}}\int d^2\sigma {{c_i}\over {1+\lambda_t-uv}}
(v\del_+u\del_-x_i-v\del_-u\del_+x_i-u\del_+v\del_-x_i+u\del_-v\del_+x_i),}
where $\lambda_i=2c_i^2/k$, $\lambda_t=\lambda_1+\lambda_2$.

As in the three dimensional case, this action can be greatly simplified
by the substitution
\eqn\redef{u=e^{\sqrt{2} t/\sqrt{k(1+\lambda_t)}}\sqrt{\hat r-(1+\lambda_t)},
v=-e^{-\sqrt{2}t/\sqrt{k(1+\lambda_t)}}\sqrt{\hat r-(1+\lambda_t)}.}
The action is then
\eqn\actfin{L={1\over {\pi}}\int d^2\sigma
(g_{\mu\nu}\del_-x^{\mu}\del_+x^{\nu}+B_{\mu\nu}
(\del_-x^{\mu}\del_+x^{\nu}-\del_+x^{\mu}\del_-x^{\nu}),}
where the metric $g_{\mu\nu}$ corresponds to the line element
\eqn\lineel{ds^2=-(1-{{1+\lambda_t}\over {\hat r}})dt^2 +(1-{{\lambda_i}\over
{\hat r}})dx_i^2+{{kd\hat r^2}\over {8\hat r^2}}[(1-{{1+\lambda_t}\over {\hat
r}})(1-{{\lambda_t}\over {\hat r}})]^{-1}
-(dx^1dx^2+dx^2dx^1){{\sqrt{\lambda_1\lambda_2}}\over {\hat r}},}
and where the antisymmetric tensor $B_{\mu\nu}$ is given by
\eqn\bmn{B_{tx_i}=\sqrt{{{\lambda_i}\over {1+\lambda_t}}}
(1-{{1+\lambda_t}\over {\hat r}}).}
\def\l{\lambda}

Henceforth, we will only consider the case where
$\lambda_1=\lambda_2=\lambda /2$, in which case,
we can diagonalize the metric by introducing
$x={1\over 2}(x_1+x_2)$ and $y={1\over 2}(x_1-x_2)$, we find the
exact same metric and antisymmetric tensor as Horne and Horowitz\hh ,
with $y$ a flat coordinate, i.e.,
\eqn\meta{ds^2=-(1-{{1+\lambda}\over {\hat r}})dt^2
+(1-{{\lambda}\over {\hat r}})dx^2+dy^2+(1-{{1+\lambda}\over {\hat r}})^{-1}
(1-{{\lambda}\over {\hat r}})^{-1} {{kd\hat r^2}\over {8\hat r^2}},}
\eqn\bmn{B_{tx}=\sqrt{{{\l}\over{1+\l}}}(1-{{1+\l}\over{\hat r}}).}

The dilaton can be calculated by considering the
determinant induced by our choice of gauge, as in \barsa , but
clearly this case will follow the three dimensional case \hh .
There, it was shown that demanding that the fields be an extremum
of the low energy effective action\callan
\eqn\leea{S=\int e^{\Phi}(R+(\nabla \Phi )^2-{1\over {12}} H^2+{8\over k}),}
(where the ${8\over k}$ cosmological constant term corresponds to the usual
$D-26$ \witta ) requires
\eqn\dila{\Phi=ln(\hat r) +a.}
Here $a$ is an arbitrary constant, which as in previous cases, will be
related to the mass of the black hole.

Since the fields are so closely related to what is found in the three
dimensional case, many but not all of the results still hold in four
dimensions.  For example, the scalar curvature is
\eqn\curv{R={{4(2\hat r +r\lambda \hat r-7\lambda-7\lambda^2)}\over
{k\hat r^2}},}
indicating that only $\hat r=0$ is a true singularity, even though the
metric components are ill-defined at $\hat r=\lambda , 1+\lambda$ as well.
As an aside, the coordinates $u,v,x,y$ are ill-defined at $uv=1$
($\hat r=\lambda $), where
our gauge fixing proceedure breaks down.  It is at this point, and
presumably in some as yet uncertain region around this point, where Eguchi
has argued that the theory is described by a topological field theory \egu .
The coordinate singularity at $\hat r =1+\lambda $ seems to have no
such interpretation.

At this point, we would like to express the fields in terms of
the mass and axionic charge of the  black hole, rather than the
arbitrary parameters $\lambda$ and $a$.  Here, we omit the details,
which follow almost immediately from the discussion in \hh , and which
yields
\eqn\qdef{Q=e^a\sqrt{{{2\lambda (1+\lambda)}\over {k}}}}
for the axionic charge per unit area, and
\eqn\mdef{M=\sqrt{2\over k}(1+\lambda)e^a}
for the mass per unit area.  Introducing the rescaled coordinate $r$ such that
\eqn\rdef{\hat r=re^{-a}\sqrt{{k\over 2}},} the final form of the fields is
\eqn\metfin{ds^2=-(1-{M\over r})dt^2+(1-{{Q^2}\over {Mr}})dx^2+dy^2
+(1-{M\over r})^{-1}(1-{{Q^2}\over {Mr}})^{-1}{{kdr^2}\over {8r^2}},}
\eqn\hfin{H_{rtx}=Q/r^2,}
\eqn\phifin{\Phi=ln(r)_+{1\over 2}ln{k\over 2}.}

\newsec{Global Structure}
The presence of two coordinate singularities reminds one of the
Reissner-Nordstr\" om metric
\eqn\rn{ds^2=-(1-{{2M}\over r}+{{Q^2}\over {r^2}})dt^2+
(1-{{2M}\over r}+{{Q^2}\over {r^2}})^{-1}dr^2+r^2d\Omega^2.}
The singularity at $r=r_+=M+\sqrt{M^2-Q^2}$ is an event horizon, while
that at $r=r_-=M-\sqrt{M^2-Q^2}$ is known as the inner horizon and is
believed to be unstable with respect to time dependant
perturbations\chandrab .  The spacetime is timelike geodesically complete.
Let us now compare these results to those for the metric in \metfin .
\medskip
\noindent{\bf A. $Q < M$}

As in the three dimensional case \hh , for the metric in \metfin ,
the singularity at $r=M$  is an event horizon similar to that at
$r=r_+$ in \rn , while the singularity at $r=Q^2/M$ is an inner
horizon similar to that at $r=r_-$ in \rn , with the only difference
being that the Killing vector $\partial /\partial t$ (timelike at
spatial infinity) becomes spacelike
from the outer horizon all the way to the singularity
at $r=0$, unlike the Reissner-Nordstr\" om case, where the Killing vector
is spacelike only between the two horizons.

There is a difference, however, between the three and four dimensional
cases when we consider geodesics, since there is an additional conserved
quantity associated with the new coordinate $y$.  Let $\xi^{\mu}$ be tangent
to an affinely parametrized geodesic, and let $E=-\xi \cdot {{\partial}\over
{\partial t}}$, $P=\xi \cdot {{\partial}\over {\partial x}}$,
$R=\xi \cdot {{\partial}\over {\partial y}}$ denote the conserved quantities.
Then the geodesics must satisfy
\eqn\geod{{{k\dot r^2}\over {8r^2}}=E^2-P^2-R^2+
{1\over r}(P^2M-{{E^2Q^2}\over M}
+R^2(M+{{Q^2}\over M}))-{{Q^2}\over {r^2}}R^2-\alpha (1-M/r)(1-Q^2/Mr),}
where the dot denotes a derivative with respect to the affine parameter, and
$\alpha$ is $0$ for null geodesics and $-1$ for timelike geodesics.  In
either case, if the right hand side of \geod\ is
positive for large $r$ then it will continue to be positive within the
horizons, and therefore geodesics
which begin at large $r$ cross both horizons.
Also in both cases, the geodesic equation \geod is eventually dominated
by a $-1/r^2$ term, so that neither null nor timelike geodesics reach
the singularity (in the three dimension case \hh , this was only true
for timelike geodesics).  Thus the spacetime is timelike and lightlike
geodesically complete.  The exact form of the geodesics will not
be needed.

The rest of the global structure, including the Penrose diagram,
is essentially identical to the three dimensional case.   For example,
the Hawking temperature can be found by analytically continuing
$t=i\tau$.  The horizon is a regular point only if $\tau$ has
period $\pi M \sqrt{2k/(M^2-Q^2)}$, corresponding to a temperature
\eqn\temp{T={1\over {\pi M}}\sqrt{{{M^2-Q^2}\over {2k}}}.}
The temperature vanishes at $Q=M$, suggesting is the black hole will
eventually settle at $Q=M$, assuming that charge cannot be radiated away.
\medskip
\noindent{\bf B. $Q=M$}

For the extremal case of $Q=M$, \metfin reads
\eqn\extrem{ds^2=(1-M/r)(-dt^2+dx^2)+dy^2+(1-M/r)^{-2}kdr^2/8r^2.}
This is similar to the extremal Reissner-Nordstr\" om metric, but
in fact, in is inappropriate to consider values of $r<M$, as
geodesics no longer go through the horizon.
The geodescis equation for this case reads
\eqn\geodext{{{k\dot r^2}\over {8r^2}}=E^2-P^2-R^2+{1\over r}(P^2M
-E^2M+2R^2M)-{{M^2R^2}\over {r^2}}+\alpha (1-M/r)^2.}
Near the (single) horizon at $r=M$, the last term is negligible, while
the remainder changes sign.  Therefore, geodesics do not penetrate the
horizon and the similarity with the Reissner-Nordstr\" om case ends.
As in \hh , it is approptiate therefore to introduce a new coordinate
$\tilde r^2=r-M$, in terms of which the metric becomes
\eqn\metqm{ds^2={{\tilde r^2}\over {\tilde r^2+m}}(-dt^2+dx^2)
+{{kd\tilde r^2}\over {2\tilde r^2}}+dy^2.}

In the new coordinate system, $\tilde r=0$ is the horizon, and geodesics
do cross the horizon.  However, the horizon does not surround a singularity,
but rather separates two identical asymptotically flat regions.
\medskip
\noindent{\bf C. $Q>M$}

A situation similar to that for $Q=M$ occurs here.  Namely, the metric
has a change of sign at $r=Q^2/M$, but this can be removed by an
appropriate choice of coordinates, as suggested by the geodesics equation.
Setting $\tilde r^=r-Q^2/M$, one finds \hh ,
\eqn\metqgm{ds^2=-{{Q^2-M^2+M\tilde r^2}\over {Q^2+M\tilde r^2}}dt^2
+{{M\tilde r^2}\over {Q^2+M\tilde r^2}}dx^2+dy^2+{{Mk}\over
{2(Q^2-M^2+M\tilde r^2)}}d\tilde r^2.}
The metric has no horizons or curvature singularities.  The conical
singularity at $\tilde r=0$ can be removed by requiring $x$ to be
periodic, which is equivalent to changing the spacetime structure at
infinity from $R^4$ to $R^3\times S^1$.  As shown in \hh , this can
also be derived from our original gauged WZW action, by gauging
the two translations together with the subgroup of $SL(2,R)$ generated
by $\pmatrix{0&1\cr -1&0\cr}$.

\newsec{Perturbation Analysis}

\def\n{\nabla}
We now proceed to consider perturbations to the dilaton, metric,
and antisymmetric tensor, following the general proceedure of
Gilbert \gnga , which follows along the lines of earlier work
of Chandrasekhar \chandraa .  The equations of motion derived
from the action \leea\ are
\eqn\anti{\n_{\lambda}H^{\lambda \mu \nu}+\n_{\lambda}\Phi
H^{\lambda \mu \nu}=0,}
\eqn\dil{-{1\over 6}H^2+\n^2\Phi +(\n\Phi )^2-{8\over k}=0,}
\eqn\grav{R_{\mu \nu}=\n_{\mu}\n_{\nu}\Phi+g_{\mu \nu}({1\over 2}
\n^2\Phi+{1\over 2}(\n\Phi )^2-{4\over k}-{{H^2}\over {12}}).}
The analysis of \gnga and \chandraa utilizes the fact that the
unperturbed metric has some symmetry, that is the components are
independant of (at least) one coordinate.  Denote the
components of the unperturbed contravariant form of the metric as
\eqn\unpert{g_{\mu \nu}=e^{2f_{\mu}}\delta_{\mu \nu},}
($\mu , \nu =0,1,2,3$).  In the present case, we will order
our coordinates as $(x^0,x^1,x^2,x^3)=(t,x,r,y)$, thus keeping
our notation as close as possible to \gnga and \chandraa .
In this case, the perturbed metric can be chosen as
\eqn\metup{g^{\mu\nu}=\pmatrix{-\fa &-\cc \fa &0&0\cr -\cc\fa &g^{11}
&\cb\fc &\cd\fd
\cr 0&\cb\fc &\fc &0\cr 0&\cd\fd &0&\fd\cr } ~}
for the contragradient form, and
\def\g{g_{\mu\nu}}
\eqn\metdn{\g =\pmatrix{(\fbp\cc^2-\fap )&-\fbp\ca &\fbp\cb\cc &\fbp\cc\cd\cr
-\fbp \cc &\fbp &-\fbp \cb &-\fbp \cd \cr \fbp \cb \cc & -\fbp \cb &
(\fbp\cb^2 +\fcp )& \fbp\cb\cd\cr \fbp\cc\cd &-\fbp\cd &
\fbp\cb\cd & (\fbp\cd^2+\fdp) \cr} ~,}
for the covariant form, where $g^{11}=\cd^2\fd+\cb^2\fc+\fb-\cc^2\fa$,
and where $\chi_0$,$\chi_2$ and $\chi_3$ (as well as all the perturbations
of $f_i$ and the other fields) are functions of $t$,$r$ and $y$ only.
Note that \metdn corresponds to a squared line element of
\eqn\ds{ds^2=-\fap dt^2+\fbp (dx-\cc dt-\cb dr-\cd dy)^2+\fcp dr^2
+\fdp dy^2 ,}
from which one can immediately determine the veirbein $e_{\mu}^a$,
\eqn\veira{e_{\mu}^a=e^{f_a},\ A=\mu ,}
\eqn\veirb{e_{\mu}^1=-\chi_{\mu}e^{f_1},\mu =0,2,3.}

As discussed in \gnga and \chandraa , the above form of the perturbed
metric has the effect of dividing the perturbations into two classes,
known as polar and axial.  Polar perturbations are those which leave
the sign of the metric unchanged upon a reversal of sign, whereas
axial perturbations are those for which one must accompany such a
reversal with the change $x\rightarrow -x$ to keep the sign of the
metric invariant.  In the present case, the variations of the $f_i$
are polar, and the $\chi_i$ are axial perturbations.  We expect
that, as in previous cases in general relativity, that the equations
for the two types of perturbations will separate.  In this paper
we will only analye the axial equations, leaving the polar perturbations
for future work.

We begin with the Einstein equation \grav .  It has been shown \chandraa
that for the axial perturbations, we only need to consider the $\{ 12\}$
and $\{ 13\}$ components of \grav .  Let us denote the right hand
side of \grav\ as $T_{\mu \nu}$, and let us use the veirbein in \veira\
and \veirb\ to go to the orthonormal frame, where Chandrasekhar has
already worked out the components of the Ricci tensor, which are
\eqn\ronetwo{R_{12}={1\over
2}e^{-2f_1-f_0-f_3}[(e^{3f_1+f_0-f_2-f_3}\chi_{23})_{,3}-
(e^{3f_1-f_0+f_3-f_2}\chi_{20})_{,0}],}
and
\eqn\ronethree{R_{13}={1\over 2}e^{-2f_1-f_0-f_2}
[(e^{3f_1+f_0-f_3-f_2}\chi_{32})_{,2}
-(e^{3f_1-f_0+f_2-f_3}\chi_{30})_{,0}],}
where $\chi_{ij}=\chi_{i,j}-\chi_{j,i}$.
Since $\chi_{ij}$ are already assumed to be first order of smallness,
when we linearize \grav , we can take $f_i$ as their unperturbed
values given by \metfin .  We now only need the form of $T_{ab}$,
again in the orthonormal frame.  In this frame, all the terms
proportional to $g_{\mu \nu }$ in \grav\  will now be proportional to
$\eta_{ab}$ and therefore will vanish because we are only considering
the cases of $a=1$, $b=2,3$. The remaining term in \grav ,
$\n_{\mu}\n_{\nu}\Phi $, must be carefully evaluated, including the
contributions of the veirbein.  The result, having eliminated terms
quadratic in small quantities, is
\eqn\dtonetwo{\delta T_{12}=0,}
\eqn\dtonethree{\delta T_{13}=-{1\over {2r}}e^{f_1-f_3-2f_2}\chi_{32}.}
The important thing to notice is that \dtonethree\ does not contain
any term proportional to $\delta \Phi $.   It can also be shown
(see below) that neither the dilaton nor antisymmetric tensor
equation (with indices raises) \grav\ and \dil\ contain terms
involving the $\chi_{ij}$.

We can now proceed to equate \ronetwo with \dtonetwo\ and \ronethree\ with
\dtonethree .  First, substitute $f_3=0$, ${k\over {8r^2}}e^{-2f_2}=
e^{2f_0+2f_1}$, and assume that all perturbations vary with
time as $e^{i\omega t}$.  Then the $\{ 12\}$ equation can be written as
\eqn\onetwo{0=e^{2f_0}\chi_{23,3}-\chi_{20,0},}
while the $\{ 13\}$ equation is
\eqn\onethree{{8\over k}e^{2f_0+2f_1}\chi_{23}= -2re^{-2f_1}({4\over
k}re^{4f_1+2f_0}\chi_{23})_{,2}-\chi_{30,0}.}
Now differentiate \onetwo with respect to $x_3$, and subtract this from
the derivative of \onethree with respect to $x_2$.  Using the fact
that $\chi_{20,03}-\chi_{30,02}=\chi_{23,00}=-\omega^2\chi_{23}$, and
defining $\alpha =re^{4f_1+2f_0}\chi_{23}$, we have
\eqn\axa{{{\del}\over {\del r}}[{8\over k}e^{-2f_1}\alpha ]= -{{\del}\over
{\del r}}[{{8r}\over k}e^{-2f_1}{{\del\alpha}\over {\del r}}]-{1\over
r}e^{-4f_1}\alpha_{,33}-{{\omega^2}\over r}e^{-4f_1-2f_0}\alpha .}
Clearly we can separate variables, and let $\alpha_{,33}=-\kappa^2\alpha$,
where $\kappa$ is some constant, which should be taken to be real so that
the solutions are well behaved at $y\rightarrow\pm\infty$.

The remaining proceedure is now clear.  First, multiply \axa by
$re^{4f_1+2f_0}$ so that the last term is just $\omega^2\alpha$.
Then make a change of variables $r\rightarrow r^*$,
so that the coefficient of ${{d^2\alpha}\over {dr^{*2}}}$ is unity.
This requires
\eqn\rstar{{{dr^*}\over {dr}}={{e^{-f_0-f_1}}\over r}\sqrt{{8\over k}}.}
Making this substitution, we have
\eqn\axb{\alpha '' +\alpha '(\sqrt{{8\over k}}e^{f_0+f_1}-f_0'-3f_1') +\alpha
(-2e^{f_0+f_1}\sqrt{{8\over k}}f_1'-\kappa^2e^{2f_0}+\omega^2),}
where the primes denote derivatives with respect to $r^*$.

Denoting by $X_1(r^*)$ the coefficient of $\alpha '$, we now eliminate
the this term by multiplying \axb by the integrating factor
\eqn\intfac{A(r^*)=e^{-{1\over 2}\int X_1(x)dx}.}
Substituting $\alpha (r^*)=A(r^*)g(r^*)$, and using the fact
that $A'(r^*)=-(1/2)X_1(r^*)A(r^*)$, etc., we have
\eqn\axc{g''+\omega^2g=V(r^*)g,}
where
\eqn\pot{V(r^*)=\kappa^2e^{2f_0}+{1\over 4}X_1^2+{1\over 2}X_1' +2\sqrt{{8\over
k}}e^{f_0+f_1}f_1'.}

This potential can easily be evaluated as a function of $r$
(implicitly a function of $r^*$ through \rstar ) by using the
explicit expressions for the $f_i$ and equation \rstar .  It is not
manifestly positive because of the $X_1'$ term.  To simplify matters
slightly, let us assume that $\kappa$ vanishes, as it only contributes
positively to the potential (for $r>M$), and thus only enhances stability.
Secondly, all the remaining terms are easily seen to be proportional to
$1/k$, so $k$ can be chosen at our convenience, say $k=8$.  Furthermore,
note that \metfin indicates that $k$ has units of
$({\rm length})^2$, so that the remaining factor
in the potential is a dimensionless function of
$r/M$ and $Q/M$.   The explicit expression for $X_1$ is then
\eqn\xone{X_1(r)=\sqrt{(1-{M\over r})(1-{{Q^2}\over {Mr}})}(1
-{M\over{2r}}(1-{M\over r})^{-1}-
{{3Q^2}\over{2Mr}}(1-{{Q^2}\over{Mr}})^{-1}).}
While the derivative of \xone\ has many terms, the only region
(outside the outer horizon) in which
the potential can turn negative is near $r=M$.  But the leading
contribution to $X_1'$ at $r=M$ goes as
\eqn\lead{X_1'(r)\rightarrow {{M^2}\over {4r^2}}\sqrt{1-{{Q^2}\over
{Mr}}}(1-{M\over r})^{-1}.}
Since the coefficient is positive, we expect the potential to be
positive near $r=M$, and therefore the potential should be positive
for all $r$.  For example, for $Q=M$, after much cancellation,
and setting $r=M/x$, the potential is simply
\eqn\meqq{V(x)={1\over 4}(1+4x-x^2).}
Since $0<x<1$, the potential is positive definite.  For more general
values of $Q/M$, some typical results are shown in Table 1.

\medskip
\noindent{{\bf Table 1:} The axial potential.}

\vbox{\offinterlineskip
\def\tablerule{\noalign{\hrule}}
\hrule
\halign{&\vrule#&
  \strut\quad\hfil#\quad\cr
height2pt&\omit&&\omit&&\omit&&\omit&&\omit&\cr
&$r/M$&&$Q=.2 M$&&$Q=.4 M$&&$Q=.6 M$&&$Q=.8 M$&\cr\tablerule
height2pt&\omit&&\omit&&\omit&&\omit&&\omit&\cr
&$2$&&$.348$&&$.365$&&$.402$&&$.484$&\cr\tablerule
height2pt&\omit&&\omit&&\omit&&\omit&&\omit&\cr
&$3$&&$.288$&&$.31$&&$.353$&&$.426$&\cr
height2pt&\omit&&\omit&&\omit&&\omit&&\omit&\cr\tablerule
height2pt&\omit&&\omit&&\omit&&\omit&&\omit&\cr
&$4$&&$.272$&&$.292$&&$.33$&&$.39$&\cr
\tablerule
&$5$&&$.265$&&$.283$&&$.315$&&$.365$&\cr\tablerule
&$6$&&$.261$&&$.277$&&$.305$&&$.348$&\cr\tablerule
&$7$&&$.259$&&$.273$&&$.298$&&$.335$&\cr\tablerule
&$8$&&$.257$&&$.27$&&$.292$&&$.325$&\cr\tablerule
&$9$&&$.256$&&$.268$&&$.288$&&$.317$&\cr\tablerule
&$10$&&$.256$&&$.266$&&$.284$&&$.311$&\cr
height2pt&\omit&&\omit&&\omit&&\omit&&\omit&\cr}
\hrule}

Is there further information which can be easily gleaned from the
coupled equations \dil -\grav \ ?  A short calculation shows that
the $(t,y)$, $(r,y)$ and $(r,t)$ components of \dil require
that $\delta H^{try}={\rm constant}$ (recall that by assumption, all
perturbations are independant of $x$).  The numerous remaining equations
all involve polar quantities such as $\delta f_i$ and $\delta \Phi$ which
will be difficult to disentangle as in the axial case.  Further
investigation of these coupled equations is currently under way.

\newsec{Discussion}
The numerical result that the axial potential is positive definite, and
that therefore the black hole is stable to axial perturbations, is
interesting in how it differs from the \lq\lq string-inspired''
case \gnga , yet incomplete.  One would like a physical explanation
for this stability,
and of course, one still must check whether
the solution is stable with respect to the polar perturbations.
Furthermore, we would like to know whether this stability is generic
for the ever increasing list of stringy black holes.  If it is not,
some criteria for stability may shed some light on the underlying
conformal field theories.  In fact, there is evidence that
it is the gauge fields, present in the string inspired models but
not in the present one, that guarantees instability, and that this
result is independant of $\alpha '$ corrections or
the presence of an axion field \gngc .  The fact that the axion
field did not enter into the equations for the axial perturbations
in the present model tends to support this evidence.

There are many other issues that can be addressed now that four
dimensional stringy black hole solutions are becoming available.
Possibly the most important of these would be to compute scattering
amplitudes, as begun in \dvv .
The fact that these models come from WZW models should
enable such calculations, and some preliminary results in the two
dimensional case are already becoming available \lyk .  Such calculations
would open the way for discussions of Hawking radiation and other
\lq\lq hair'' which a black hole may have, as well as the role of
string effects such as winding states.

Another intersting question which was mentioned briefly in Section 2 is
the recent result of Eguchi \egu\ that the horizon in the two
dimensional black hole is described by a topological field theory.
The first question such a result raises is what region {\it around}
the horizon is described by such a theory.  As topological field
theories are by construction independant of the metric, it is not
clear how any length scale which determines this region can enter
the calculation.  Assuming the generality of this result, however,
we see that in the four dimensional case, the outer horizon, which is
stable to perturbations, is described by a topological field theory,
whereas the inner horizon, which we expect to be unstable
by the analogy with the Reissner-Nordstr\" om case, is not.  It is
tempting indeed to conjecture that this correlation is in fact general.

Yet another avenue for further investigation concerns string loop
contributions.  In a very recent paper, Ellis et.al. \ellis argue
that the $c=1$ matrix model represents the evaporation endpoint
of the two dimensional black hole due to the absence in the matrix
model of imaginary parts in higher genus amplitudes.  The arguments
do not depend on the number of background dimensions, and therefore
appear to be general.  They further argue that the static nature
of the known black hole solutions will be destroyed at higher
genus.  If this is true, detailed calculations of the evaporation
process and a comparison to conventional black holes would be
important to pursue.

In summary, it would seem that there are many interrelated
issues involving string theory and black holes, of which only
a few have even begun to be addressed.  Hopefully, further
analysis of these toy models will shed significant light
on the important underlying physics.
\vfill
\eject
\listrefs
\bye